\documentclass[usenatbib]{mn2e}
\usepackage{deluxetable}
\usepackage{graphicx}
\usepackage{lscape}
\newcommand{\Ly}{\mbox{Ly$\alpha$}}
\newcommand{\NV}{N{\sc v}}

\newcommand{\CIV}{C{\sc iv}}
\newcommand{\OI}{O{\sc i}}
\newcommand{\HB}{\mbox{H$\beta$}}
\newcommand{\CIIIf}{C{\sc iii}]}
\newcommand{\MgII}{Mg{\sc ii}}

\def\lesssim{\mathrel{\hbox{\rlap{\hbox{\lower3pt\hbox{$\sim$}}}\hbox{\raise2pt\hbox{$<$}}}}}

\title[Quasar Variability]{
Variability of Broad Emission Lines in High-Luminosity, High-Redshift Quasars}
\author[S. C. Woo et al.]
{\parbox[t]{\textwidth}{\raggedright Sui Chi Woo$^{1}$
\thanks{E-mail: suw11@pitt.edu}, 
David A. Turnshek$^{1}$, 
Carles Badenes$^{1}$, 
and 
Steven Bickerton$^{2}$}
\vspace*{6pt}\\
$^{1}$PITTsburgh Partical physics, Astrophysics and Cosmology Center (PITT PACC), 
Department of Physics and Astronomy,\\
University of Pittsburgh, Pittsburgh, PA 15260, USA\\
$^{2}$Department of Astrophysical Sciences, Princeton University, 
Princeton NJ, USA
}

\begin{document}

\date{}

\pagerange{\pageref{firstpage}--\pageref{lastpage}} \pubyear{2011}

\maketitle

\label{firstpage}

\begin{abstract}

We examine the variability of the high-ionization \Ly\ $\lambda 1216$ broad emission
line (BEL) in a sample of 61 high-luminosity, high-redshift quasars 
observed at two epochs by the Sloan Digital Sky Survey. These bright objects 
lie in the redshift interval $z=[2.5, 4.3]$ 
and have luminosities $3.4 \times 10^{45} \lesssim \lambda L_{\lambda}
\lesssim 3.4 \times 10^{46}$ 
${\rm erg\,s^{-1}}$ at 1450 \AA. 
Utilizing improved spectrophotometric flux calibrations relative to nearby compact stars observed 
simultaneously, we are able to measure the flux changes 
in \Ly\ and the nearby continuum at 
two epochs. 
We find 20 objects that exhibit \Ly\ BEL flux variability at a significance level  
greater than $5\sigma$ on time-scales ranging from 
days to years in the quasar rest frame. The results show that, 
although some earlier work showed no significant detections
of \Ly\ BEL flux changes in a quasar sample with even higher luminosity,
variability is present and readily observable in the sample studied here. 
We also consider the \CIV\ $\lambda1549$ BEL. The lack of a strong correlation
between \Ly\ BEL variability and nearby continuum variability is consistent with the 
presence of a time lag between the variations, whereas the presence of a stronger correlation between \Ly\ BEL
variability and \CIV\ BEL variability suggests that these BEL regions (BELRs) are at similar distances
from the central ionizing source. Some interesting examples are high-lighted in the analysis,
including a case where the flux of an \Ly\ BEL increased by
$\sim26$\% in 14 days in the quasar rest frame, suggesting that 
the BELR has the shape of a disc, which is being observed face-on. 
This work demonstrates that future campaigns of 
spectrophotometric monitoring can improve our understanding 
of the structure of the BELRs of high-luminosity, high-redshift quasars.

\end{abstract}


\section{Introduction}

Broad emission lines (BELs), such as \Ly, \NV\ and \CIV, are 
prominent signatures seen in the optical/UV spectra of quasars. 
These lines arise in gas photoionized by the X-ray and UV continuum emission 
from the inner region of an accretion disc surrounding 
a supermassive black hole (SMBH) \citep{pet11,pet06,elv10,dwb07}. 
Consequently, when continuum fluxes vary 
with time, BEL fluxes should respond to these  
variations but with time lags. Reverberation mapping studies 
of individual quasars and active galactic nuclei (AGN)
can use results on time lags to determine the distances between 
central continuum sources and BEL regions (BELRs). 
Furthermore, these
size estimates for BELRs can 
be used with BEL velocity widths to infer the masses of 
SMBHs, under the assumption that the line-emitting regions 
are gravitationally bound to the SMBHs \citep{gpp+12,dpp+10,vp06,kmn+05}. 
Deducing BELR size scales and SMBH masses 
are fundamental to understanding the physics 
of quasars.  
 
There have been numerous reverberation mapping studies 
of lower luminosity, low-redshift AGN 
such as Seyfert galaxies. See, for example, \citet{pet93} and \citet{pfg+04} for early reviews. 
Observationally such objects are brighter because they are nearby, and therefore variations are 
easier to detect. Also, their BELs tend to vary on shorter 
time-scales, of the order if light days to light weeks, and this permits more convenient 
temporal sampling to study the variability. 
Studies of lower luminosity AGN show that  
higher ionization BELs (e.g., \Ly\ and \CIV) 
respond more rapidly to continuum variations than 
lower ionization BELs (e.g., \HB, \CIIIf, and \MgII), 
indicating that lower ionization BELRs are farther out from 
the central source. In particular, the eight-month UV monitoring programme 
of NGC 5548 ($\lambda L_{\lambda} \sim 4.2 \times 10^{43}$  ${\rm erg\,s^{-1}}$ at 1300 \AA)
found time lags of $\sim10$ days for both \Ly\ and \CIV, suggesting that  
these BELRs have similar size scales \citep{cra+91}.  

To date, limited reverberation mapping work has been attempted on the more  
luminous quasars. Although quasars are known 
to be more energetic than AGN, they are usually significantly fainter than well-studied AGN due to their
higher redshifts. Moreover, under the assumption that the photoionized BELRs of quasars and AGN have 
similar physical conditions, quasar BELR distances from the central source (and hence 
time lags) are expected to be larger 
by a factor of $\sim (L_{quasar}/L_{AGN})^{0.5}$, where $L$ is luminosity. This is exacerbated by 
the fact that time lags in observed frames should be a factor of $1+z$ larger than in quasar/AGN rest frames
due to cosmological considerations (i.e., at high redshifts we observe variations in ``slow motion'').
All of these effects make reverberation mapping 
studies of high-luminosity, high-redshift quasars more difficult and challenging.

While numerous studies have
demonstrated that quasars have variable continua, 
few studies of their   
BEL variability have been made. 
\citet{oh91} found a time lag of $\sim 74$ days between the UV continuum and 
the \Ly\ BEL in the luminous quasar 3C273.   
However, \citet{ucw93} argued that the BEL variations were only 
marginally significant. A number of years ago \citet{kbm+07} reported no 
significant (i.e., $>5$\%) \Ly\ BEL variability in a very
luminous sample of high-redshift quasars.
\footnote{The sample studied here is not as
luminous. The luminosities of our sample at 1450\AA after galactic extinction corrections 
range from $3.40\times10^{45}$ to $3.40\times10^{46}$ ${\rm erg\,s^{-1}}$, 
while their sample has luminosities from $2.76\times10^{46}$ to 
$4.27\times10^{47}$ ${\rm erg\,s^{-1}}$.} 
However, they did report significant variations in the \CIV\ BEL 
in response to continuum variations, leading to 
a time lag estimate of $\sim 188$ days in the quasar rest frame. 
Given these ambiguous results, it is not clear whether 
the \Ly\ BEL shows the same variability results as the \CIV\ BEL, and whether
reverberation mapping results from low-luminosity, low-redshift AGN can be scaled and applied to  
high-luminosity, high-redshift quasars.

Therefore, the goal of this paper is to investigate the variability of  
BELs in high-luminosity, high-redshift quasars. This would help 
clarify the feasibility of reverberation 
mapping studies for such a sample. To accomplish this we 
selected a sample from the Sloan Digital Sky Survey 
(SDSS) which had multiple spectral observations 
and we applied the spectrophotometric calibrations 
of \citet{ycv+09} to improve the spectrophotometry. 
In Section~\ref{sec:obs} we present the quasar sample, 
and we discuss the additional calibrations which can improve the relative
spectrophotometric accuracy of the data. 
In Section~\ref{sec:analysis} we outline how we measure
flux changes in BELs and continua, and we present the results.
Among other findings, our analyses show that $\sim$33\% of our sample exhibits 
significant ($>5\sigma$) \Ly\ BEL flux variations. These variations are not 
strongly correlated with nearby continuum variations, which is consistent
with the presence of a time lag. However, a stronger statistical 
correlation between \Ly\ BEL flux variations and \CIV\ BEL flux variations is seen,
which suggests that these BELRs have similar size scales.
In Section~\ref{sec:concl} we summarize the results and briefly discuss future prospects.

\section{Quasar Data Set}
\label{sec:obs}
\subsection{Sample Definition}
For this study we used the SDSS data release (DR8, \citet{aaa+11}) to 
select an appropriate sample of quasars. In the observed frame
SDSS spectra cover the optical to near-infrared 
($3800-9200$ \AA) with resolution $\lambda/\Delta\lambda\sim2000$ 
at 5000 \AA\ \citep{slb+02}. To define our quasar sample and 
ensure that \Ly\ $\lambda1216$ BELs are included in the spectra, we selected objects 
which are spectroscopically classified as QSOs, have redshifts in the interval $z=[2.5 - 5.0]$,  and have 
multiple spectroscopic observations. 
The spectral classifications given in DR8 
are highly reliable, with 98\% accuracy 
in two independent reduction pipelines (spectro1d and idlspec2d). 
We excluded quasars observed using different plates on different dates to avoid possible 
calibration problems. 
Reasons for repeat spectroscopic observations in the SDSS 
are described in \citet{wvk+05}. 
The repeat observations span 
time intervals ranging from days to years in the observed frame, leading to 
interesting durations for probing spectral variability in the quasar rest frames. 
In addition to the wide range of time intervals, 
the large sample size and homogeneity of SDSS spectroscopic data 
provide a unique advantage. 

We found 407 quasars meeting the above criteria in DR8. 
Since variability in the observed frame at these higher redshifts is generally expected to increase with
time interval, we have used only the first and last observations in order to 
maximize the time interval when more than two observations at different dates with the same plate exist. 

\subsection{Refinement of the Spectrophotometric Calibration}\label{sec:cal}
The work of \citet{vwk+04} and \citet{wvk+05} demonstrates that 
some additional spectrophotometric calibrations of 
SDSS spectra can be applied to increase the accuracy 
of spectral variability studies. \citet{wvk+05} used 
the stars to derive calibration differences 
between same-plate pairs, under the assumption that the majority 
of stars are non-variable. 
Later \citet{ycv+09} adopted a slightly different methodology, 
using galaxies instead of stars, since galaxies should be non-variable. 
\citet{ycv+09} estimated that their method improves the relative flux calibration between
plate pairs by about a factor of two. 
The two-step calibration methods we used in this study are summarized below. See 
\citet{ycv+09} for a complete discussion. 

First, to achieve an improved relative flux calibration for objects 
on a pair of plates taken at different modified Julian dates (MJDs), 
a wavelength-dependent flux ratio correction calibration is derived 
and applied. This is step one. This is done by employing corrections made available
by \citet{ycv+09}. Their final corrections were derived from fifth-order
polynomial least-squares fits, which provide low-order spectral corrections and eliminate higher-order 
spurious spectral features. These corrections were available 
for plate pairs in our sample up to DR6. 
Therefore, this further limited our
sample size, since only 181 of the 407 plate pairs in our DR8 sample had an available 
wavelength-dependent calibration correction. This correction was applied to all 181 objects
to search for variability.  

Secondly, for significantly variable quasars, which are identified after step one and which require closer
examination, 
we can derive and apply a wavelength-independent (i.e., grey)
flux normalization correction using measurements of nearby non-variable compact objects surrounding a 
quasar.\footnote{A better approach might involve using galaxies to further improve the flux calibration, since they are
clearly non-variable. However, because galaxies are extended and quasars are compact, possible variations in 
seeing between plate epochs make this approach problematic.}
This improves the relative spectrophotometric calibration between plates. 
We call this correction $C_{band}$, where ``band'' refers to the rest
wavelength intervals specified below. Instead of choosing a narrow wavelength band at 
6450 \AA\ $\pm$ 520 ${\rm km\,s^{-1}}$ as suggested in \citet{ycv+09}, 
we use the entire region from just longward of \Ly\ to longward of \CIV, 
but exclude the locations of prominent emission lines. 
The rest frame wavelength bands which we use to derive $C_{band}$ are  
1260--1286, 1318--1380, 1430--1525, and 1580--1620 \AA.\footnote{The 
derived $C_{band}$ results for each compact object 
change less than a few percent when the wavelength intervals are reduced by a factor of two to 
1260--1273, 1318--1349, 1430--1478, and 1580--1600 \AA. 
Thus, derived $C_{band}$ corrections are robust and reasonably optimized.} 
Table~\ref{table1} gives examples of the additional improvements that can be achieved by presenting
before and after $C_{band}$ corrections for
four selected quasars and nearby non-variable compact objects. These four example
quasars are further discussed in Section~\ref{sec:eg}.
After $C_{band}$ corrections are applied, compact objects should show no significant
variability, and this turns out to be the case. 

In summary, as outlined above, we implemented the first step 
to improve the wavelength-dependent spectrophotometric calibration 
and then the second step to improve the wavelength-independent relative spectrophotometric
calibration, before we performed the variability analysis in Section~\ref{sec:analysis}.

\subsection{Elimination of Problematic Objects}

\subsubsection{Poor Sky Subtraction}
\begin{figure*}
\centering
\includegraphics[width=0.99\textwidth,angle=0]{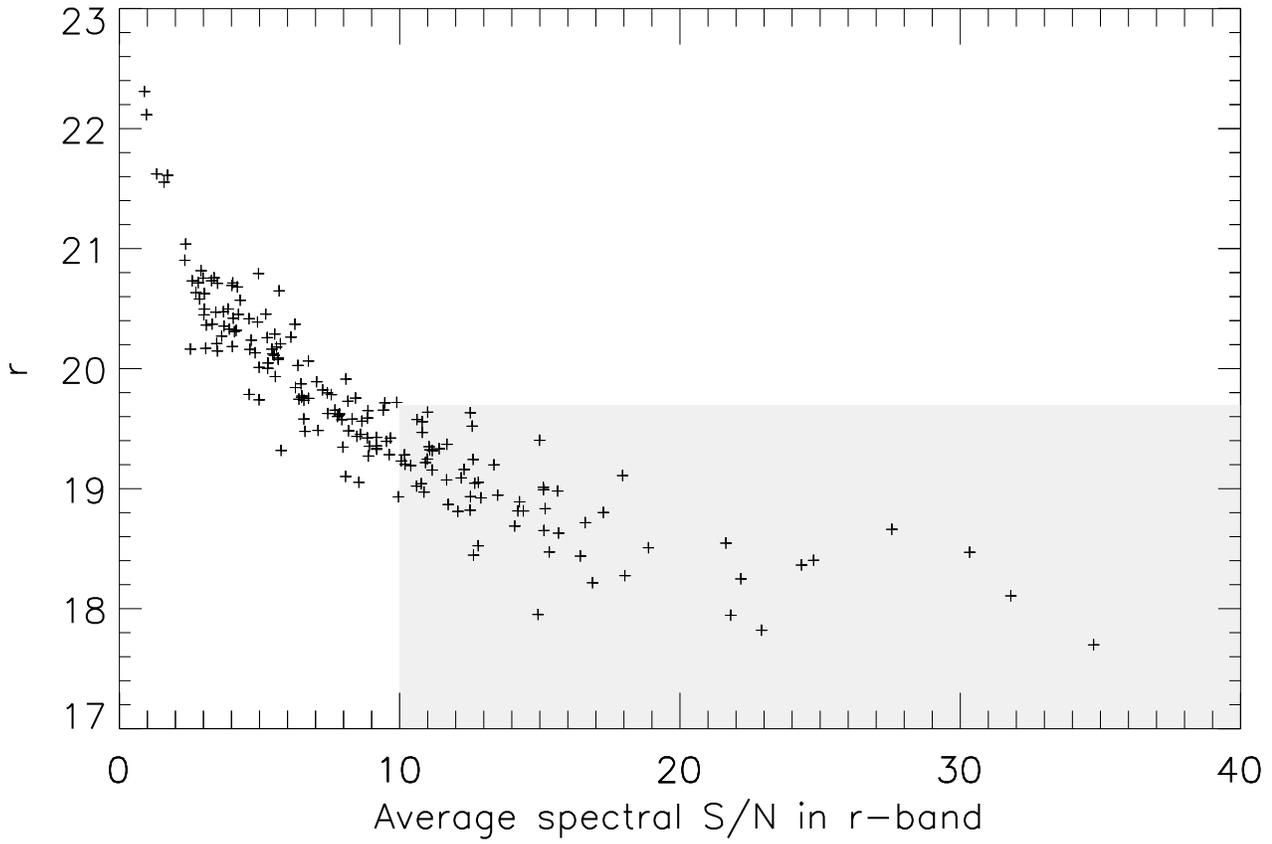}
\caption{Average SDSS $r-$band magnitude measured from the spectra at two
epochs versus average spectral S/N. 
Objects with S/N $<$ 10 and $r$ fainter than $\sim19.7$ mag are eliminated from the final sample to 
avoid possible false-positive results on variability due to sky subtraction
residuals.}
\label{PlotMagrVsSN_bfinal}
\end{figure*}

\begin{figure*}
\centering
\includegraphics[width=0.99\textwidth,angle=0]{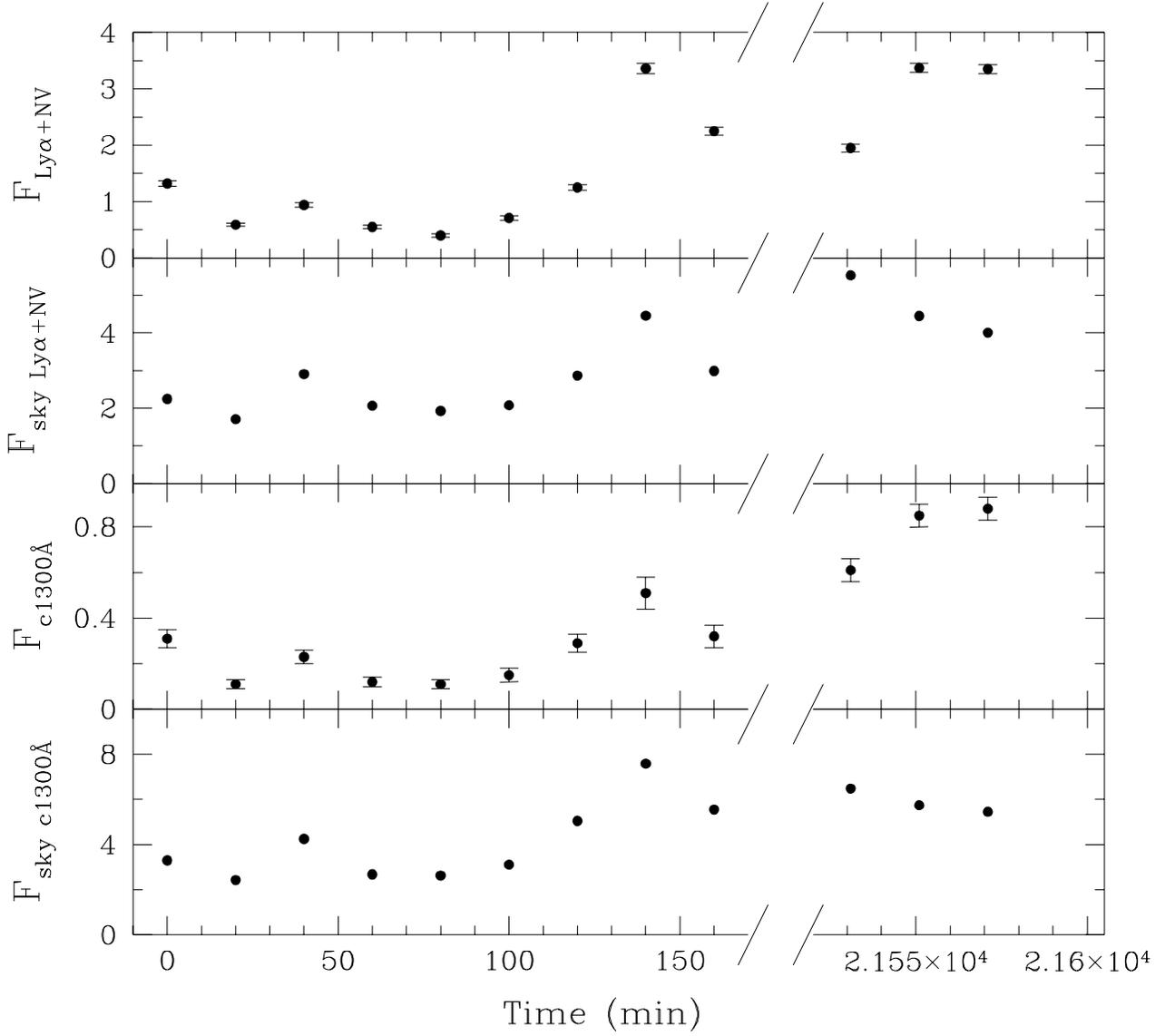}
\caption{Faint quasar SDSS J101840.46+285000.7 with flux variations  
correlated to sky fluctuations in \Ly\ $+$ \NV\ BEL regions (top two)
and nearby continuum regions (bottom two), respectively, 
over 12 sub-exposures. 
The unit of flux is $10^{-17}$\,erg\,s$^{-1}$\,cm$^{-2}$\,\AA$^{-1}$.}
\label{qsosky}
\end{figure*}

Sky subtraction of SDSS spectra is done in an automated data reduction pipeline. 
\citet{wh10} have pointed out that significant systematic residuals can remain in
SDSS sky-subtracted spectra due to sub-optimal subtraction of the strong OH sky emission lines longward
of 6700 \AA. In addition, we have found that some lower signal-to-noise ratio (S/N) spectra 
(e.g., spectra of fainter high-redshift quasars initially 
selected for this study) can suffer from poor sky subtraction, leading to 
false-positive results on spectral variability. For the two observation epochs of our sample of 181 quasars,
we plot in Figure~\ref{PlotMagrVsSN_bfinal} the average synthetic
 r-band magnitude derived from the SDSS spectra 
versus the average spectral S/N.
As expected, S/N is seen to decrease with increasing magnitude. However, for average $r> 19.7$ or
average S/N $<$ 10 we have identified cases of false-positive results when performing our variability analysis.
A clear indication of potential problems can be seen by examining
data for the faint quasar SDSS J101840.46+285000.7, which has 
average $r \approx 22.3$ and average spectral S/N $\approx$ 0.9. 
In our initial analysis of this object we detected significant flux variations 
over the entire range of observed wavelengths in just $\sim2.5$ days in 
the quasar rest frame. However, examination of the sub-exposures \citep{bbh+12}, 
which were used to form the SDSS spectra at the two epochs, showed that poor sky 
subtraction was responsible for the apparent flux variations. The initial epoch spectrum
was formed from nine sub-exposures and the final epoch spectrum was formed from three
sub-exposures. 
Figure~\ref{qsosky} 
illustrates the correlations between fluctuating sky spectra and 
\Ly\ $+$ \NV\ BEL fluxes and nearby continuum fluxes (see Section~\ref{sec:consub} 
for how these were measured), respectively, over 12 sub-exposures.\footnote{
Initially we considered the \Ly\ $+$ \NV\ BEL blend, 
but later we focus on the \Ly\ BEL. For the analysis presented in Figure~\ref{qsosky}, 
we measured the \Ly\ $+$ \NV\ BELs, but for the subsequent analyses 
we measured the \Ly\ BEL.} 
Following the initial analysis of our sample we have concluded that a reliable 
search for spectral variability 
requires the exclusion of objects with average $r-$band magnitudes 
fainter than $\sim 19.7$ and average spectral S/N $<$ 10. 
This reduces our sample to 64 quasars. 

\subsubsection{Discrepant Redshifts}
When considering quasar redshift information for our sample 
we had to exclude three objects because the SDSS pipeline redshifts
were discrepant: SDSS J101449.56+032600.8, SDSS J105905.06+272755.4,
and SDSS J115852.86$-$004301.9.\footnote{
For SDSS J101449.56+032600.8 and SDSS J105905.06+272755.4, 
the BELs are incorrectly identified so the redshifts are incorrect. 
For SDSS J115852.86$-$004301.9, the classification of QSO is incorrect.} 
This reduces the size of our final sample to 61
quasars with luminosities
$\lambda L_{\lambda} > 3.4\times10^{45}$  ${\rm erg\,s^{-1}}$ at 1450 \AA\ and redshifts
$z=[2.5, 4.3]$. 

\section{Variability Analysis}
\label{sec:analysis}
Due to S/N differences and velocity offsets among different measured BELs, 
the SDSS pipeline usually calculates 
slightly different emission redshifts for the pair of spectra being analysed. 
Also, \citet{wh10} have shown that SDSS quasar
redshifts possess systematic biases of $\sim600$ km s$^{-1}$.
For consistency, and because of the effects mentioned above, we take two precautions
when performing our analysis. First, we calculate a single redshift by averaging the pair of redshifts
available for the two spectra. We then shift each spectrum
into the rest frame of the quasar for subsequent analysis. 
The rest-frame wavelength region of interest was chosen to be 
1100--1700 \AA\ so that the regions containing \Ly\ $\lambda1216$ BELs and 
\CIV\ $\lambda1549$ BELs are included. 
Secondly, as discussed below, when measuring BEL fluxes, we use a 
velocity window which is large enough to ensure that typical velocity offsets will not significantly affect
the results.  

As discussed in the next two sections, we use two methods to make two different kinds of comparison to
assess variability. 
Both methods rely on determining and subtracting the continua underneath 
the BELs in each spectrum. 
In the first method flux differences of BELs and continua between the two spectra are considered. 
In the second method, flux ratios of
BELs and continua between the two spectra are considered. 

\subsection{Continuum and BEL Flux Changes}\label{sec:consub}

\subsubsection{Continuum Changes}
\label{subsec:cc}
For each two-epoch pair of spectra 
we start by determining the mean continuum flux of each spectrum 
in the line-free regions near the \Ly\ BEL, defined as 1276--1286 \AA\ and
1318--1334 \AA, 
which ensures that the region in the vicinity of any \OI\ $\lambda1302$ BEL is excluded. 
Also, if necessary, any bad spectral regions within these wavelength intervals are identified
by visual inspection and excluded. 
The spectrum with higher S/N in the observed r-band
in the two-epoch pair is labeled HSN, 
whereas LSN refers to the one with lower S/N. 
The continuum flux difference at $\sim1300$ \AA\ is determined by
subtracting the LSN-epoch mean flux 
from the HSN-epoch mean flux: 
$\Delta F_{c1300}
=F_{c1300,HSN}-F_{c1300,LSN}$. 
The normalized flux change in the continuum 
is then defined to be
$\Delta F_{c1300}/{F_{c1300, HSN}}$. Since the change in continuum level near $\sim1300$ \AA\ is
as near as we can get to the ionizing continuum, we use this wavelength to measure the continuum change throughout
this study. Standard deviations are 
determined by propagating the SDSS pipeline flux errors. 
Twenty-nine objects exhibit continuum flux changes at levels of significance $>5\sigma$.

\subsubsection{\Ly\ BEL Flux Changes}
\label{subsec:lybel}
To study flux variations in the \Ly\ BEL
we chose 1204--1229 \AA\ as the region of interest (i.e., a region encompassing 
$\approx 6170$ km s$^{-1}$ around the BEL). This  
generally covers most of the flux in a \Ly\ BEL and is selected to 
mostly avoid inclusion of the nearby \NV\ $\lambda1240$ BEL. Thus, the chosen interval is a practical fixed choice
not intended to match the observed width of the \Ly\ BEL, but it is suitable for the purpose of this study. 
The measured change in BEL flux can weakly depend on the chosen wavelength 
interval.\footnote{Also, a flux change might vary across the BEL profile. For example, one or both of the BEL 
wings could show changes that are different than changes in the BEL core.  We searched for clear 
indications of such behaviour by considering the 
central half of the original velocity interval  (i.e., 1210--1223 \AA), but found no clear examples 
of this. However, see Section~\ref{sec:eg_no3}.\label{fn:4}} 
To determine the continuum level to subtract underneath a \Ly\ BEL we use
the mean continuum flux in the nearby continuum (Section~\ref{subsec:cc}). This works as well as an extrapolated 
straight-line fit. An extrapolated higher-order 
polynomial fit would be inappropriate. Unfortunately, the region shortward of the \Ly\ BEL is generally
depressed by the \Ly\ forest in the moderate resolution SDSS spectra, so this region could 
not be used to help determine the continuum level underneath the \Ly\ BEL.
The absolute \Ly\ BEL fluxes in the HSN and LSN pair of spectra are then $F_{\Ly,HSN}$ and $F_{\Ly,LSN}$, 
respectively. Consistent with our previous definitions for a change in the continuum, the \Ly\ BEL 
flux difference is 
$\Delta F_{\Ly}=F_{\Ly,HSN}-F_{\Ly,LSN}$ and the normalized \Ly\ BEL flux difference is
$\Delta F_{\Ly}/{F_{\Ly, HSN}}$.

\begin{figure*}
\centering
\includegraphics[width=0.8\textwidth,angle=0]{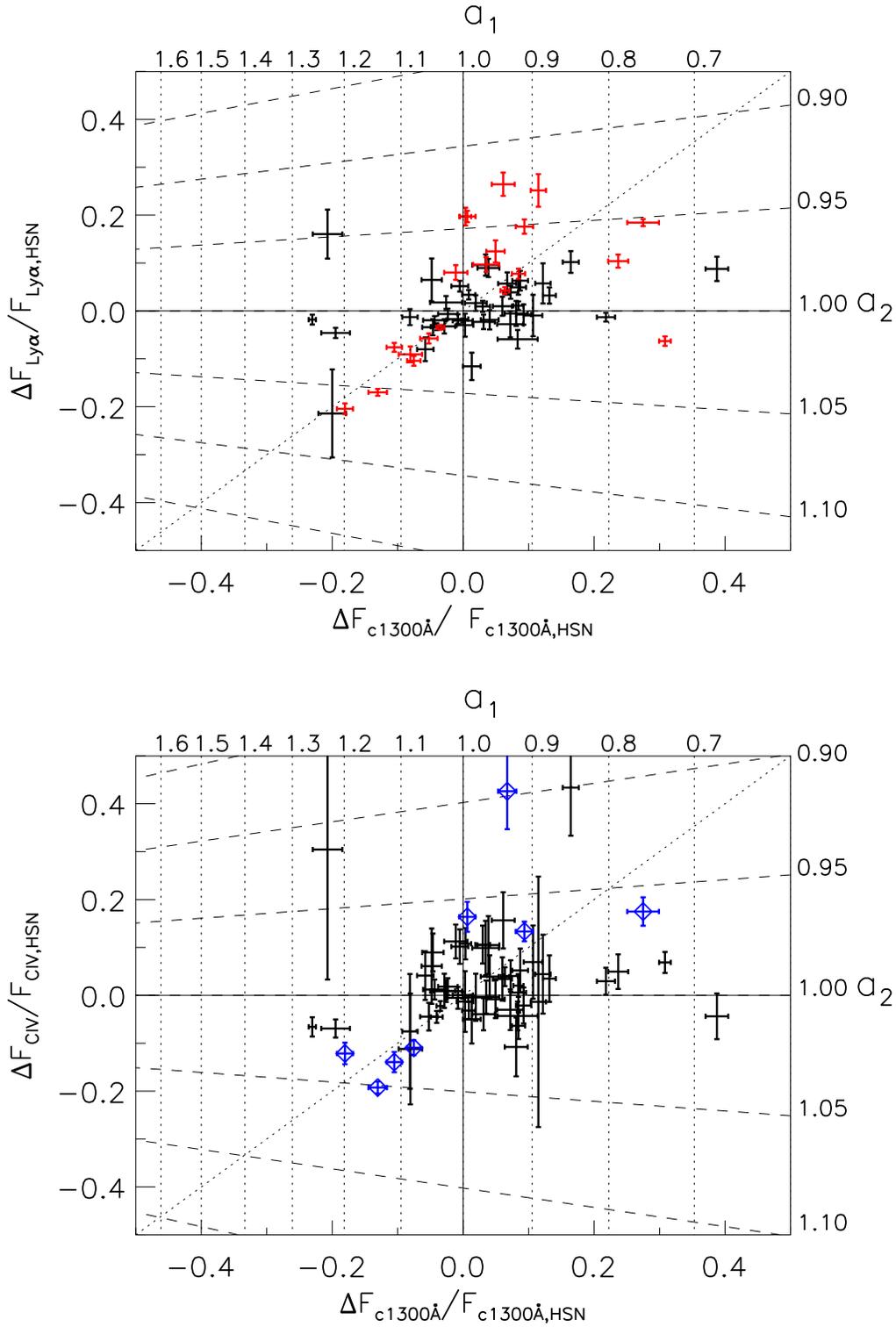}
\caption{Top: The distribution of normalized \Ly\ BEL flux 
changes vs. normalized continuum flux changes for 61 quasars. 
Twenty quasars (red points) exhibit \Ly\ BEL flux variations 
at a significance level $>5\sigma$. 
Bottom: The distribution of normalized \CIV\ BEL flux changes 
vs. normalized continuum flux changes for 61 quasars. 
Eight quasars (blue diamonds) exhibit \CIV\ BEL flux variations 
at a significance level $>5\sigma$. 
We have parameterized flux changes in a hypothetical \Ly\ BEL profile 
and the associated nearby continuum in the parameter space of $a_1$ 
and $a_2$. See the details in the text.}  
\label{P_LyACIVVsLyAc_bfinal}
\end{figure*}

Figure~\ref{P_LyACIVVsLyAc_bfinal} shows the distribution of normalized \Ly\ BEL flux changes
versus normalized continuum flux changes for the 61 sample quasars.
Again, the standard deviations shown for the plotted points are 
determined by propagating the SDSS pipeline flux errors.
Twenty quasars (shown in red) show \Ly\ BEL 
flux variations at a significance level 
$>5\sigma$.
Among these 20 quasars we have selected four
to discuss in Section~\ref{sec:eg}. 
 
As an illustration of what the results shown in Figure~\ref{P_LyACIVVsLyAc_bfinal} 
might mean we have parametrized how 
flux changes in a hypothetical, but typical, \Ly\ BEL profile and the associated nearby continuum 
would appear in this plot. To do this 
we parametrize a continuum-normalized HSN spectrum 
as 
\begin{eqnarray}
f_{HSN}(\lambda)=1+Ae^{-(\frac{\lambda-1216\mathring{\rm A}}
{\delta\lambda})^2},
\end{eqnarray}
where $A$ is the amplitude of the \Ly\ BEL peak at 1216 \AA\ relative to 
the normalized continuum and $\delta\lambda$ sets the line width.
For this illustration we put $A=5.5$ and $\delta\lambda = 12$ \AA\ ($\sim3000$ km s$^{-1}$), which are typical.
The hypothetical LSN spectrum is then paramertrized as
\begin{eqnarray}
f_{LSN}(\lambda)=a_1+a_2Ae^{-(\frac{\lambda-1216\mathring{\rm A}}
{\delta\lambda})^2}, 
\end{eqnarray}
where $a_1$ and $a_2$ dictate 
how the flux in the continuum and \Ly\ BEL change, respectively.
A grid of $a_1$ (on the dashed horizontal axis) and $a_2$ (on the dotted vertical axis) parameters 
are overplotted on Figure~\ref{P_LyACIVVsLyAc_bfinal} to give
an indication of the nature of the observed flux changes in our sample.

\subsubsection{\CIV\ BEL Flux Changes}
Following the procedure we used to measure \Ly\ BELs in Section~\ref{subsec:lybel}, 
we have also measured \CIV\ BELs. However,
the continuum levels we subtract from \CIV\ BELs are measured at 1470--1488 \AA\ 
and 1600--1616 \AA\ in the quasar rest-frames. 
We then measure the \CIV\ BEL over the rest-frame interval 1534--1566 \AA, which corresponds to 
a velocity interval that is similar to the one used for the \Ly\ BEL. 
Eight objects exhibit (shown in blue) \CIV\ BEL flux changes at levels of significance $>5\sigma$.
In the bottom panel of Figure~\ref{P_LyACIVVsLyAc_bfinal} we  
show $\Delta F_{CIV}/{F_{CIV, HSN}}$ versus 
$\Delta F_{c1300}/{F_{c1300, HSN}}$. 
On this bottom panel we also show the same 
grid of $a_1$ and $a_2$ values shown in the top panel of Figure~\ref{P_LyACIVVsLyAc_bfinal}
for the \Ly\ BELs. 
This permits a comparison between changes in the 
\Ly\ and \CIV\ BELs. 

\subsubsection{Correlations between Flux Changes in the Continuum, \Ly\ BEL, and \CIV\ BEL}
\label{subsec:corrF}
\begin{figure*}
\centering
\includegraphics[width=1\textwidth,angle=0]{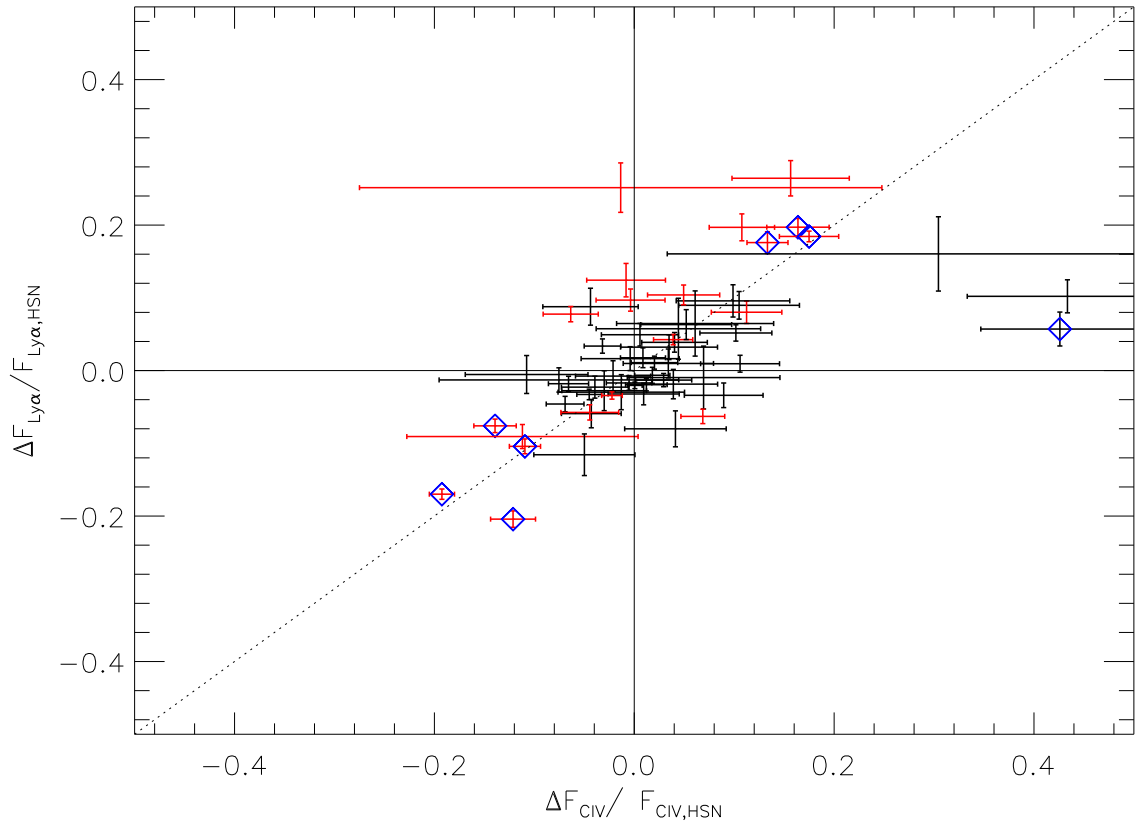}
\caption{Fractional change in \Ly\ BEL flux vs. 
fractional change in \CIV\ BEL flux between two epochs.} 
\label{P_LyAVsCIV_bfinal}
\end{figure*}

Past monitoring campaigns of low-luminosity Seyfert 1s  
conducted with the {\it International Ultraviolet Explorer} 
have revealed similar time lags for \Ly\ and \CIV\ BELs \citep{kab+95}. 
To check if this holds in our quasar sample, 
in Figure~\ref{P_LyAVsCIV_bfinal} we plot the normalized \Ly\ BEL flux changes versus the
normalized \CIV\ BEL flux changes.
Analysis of the results shown in Figures~\ref{P_LyACIVVsLyAc_bfinal} and \ref{P_LyAVsCIV_bfinal}
reveal correlations between the flux change results. 
After taking into account the variances of the data points, the correlation coefficients are: 
0.43 for \Ly\ BEL flux changes versus continuum changes (Figure~\ref{P_LyACIVVsLyAc_bfinal}, top panel), 
0.51 for \CIV\ BEL flux changes versus continuum changes (Figure~\ref{P_LyACIVVsLyAc_bfinal}, bottom panel), 
and 0.74 for \Ly\ BEL flux changes versus \CIV\ BEL flux changes (Figure~\ref{P_LyAVsCIV_bfinal}). 
Thus, there is a significantly stronger correlation between observed changes in the two BELs, 
in comparison to correlations between changes in either BEL and the continuum.   

\subsection{Flux Ratios between the HSN and LSN Epochs}
\begin{figure*}
\centering
\includegraphics[width=0.8\textwidth,angle=0]{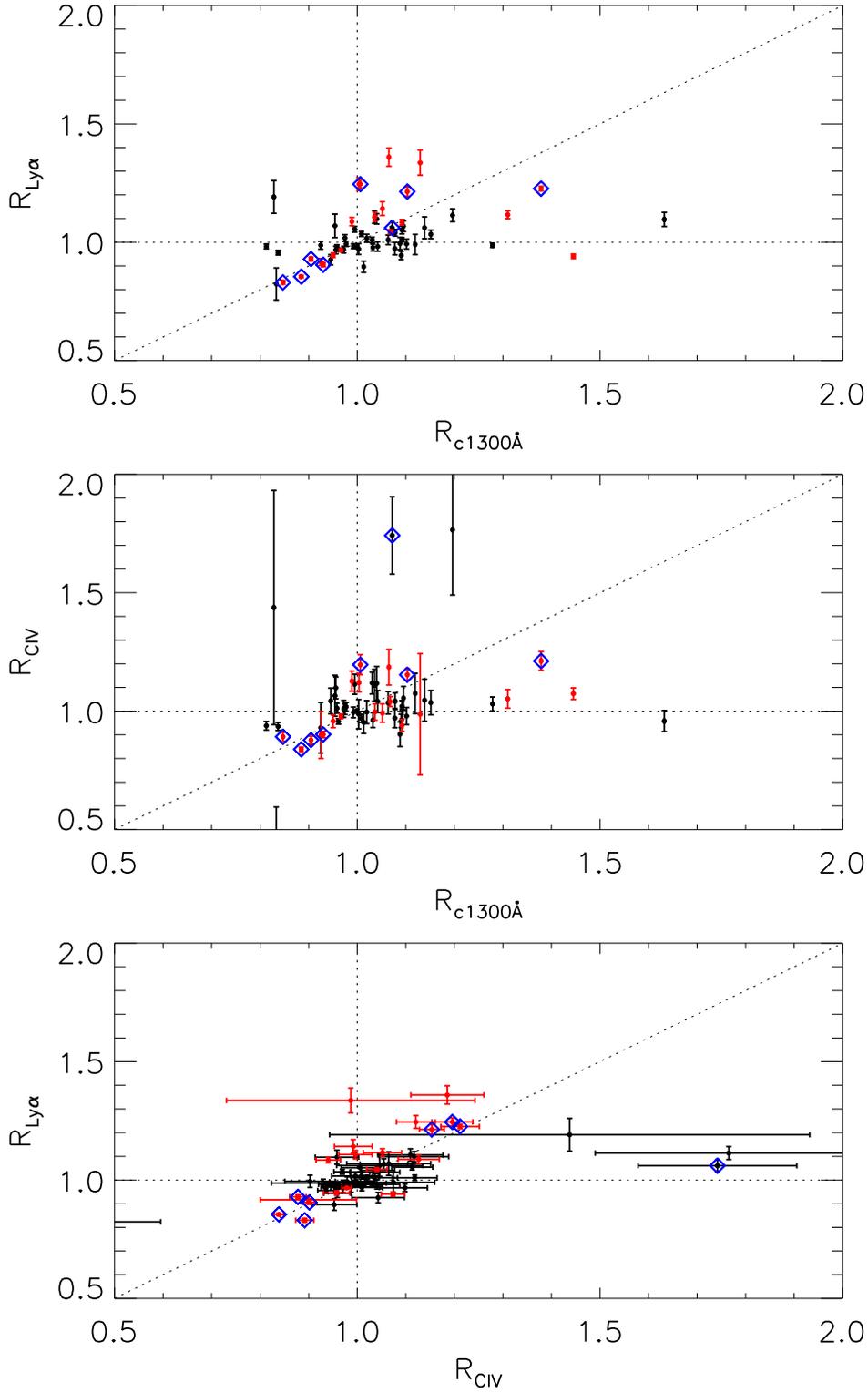}
\caption{Top: \Ly\ BEL flux ratio vs. continuum flux ratio. 
Middle: \CIV\ BEL flux ratio vs. continuum flux ratio.
Bottom: \Ly\ BEL flux ratio vs. \CIV\ BEL flux ratio. 
As in Figures~\ref{P_LyACIVVsLyAc_bfinal} and \ref{P_LyAVsCIV_bfinal}, 
the red points and blue diamonds are those with \Ly\ and \CIV\ BEL flux
changes $>5\sigma$, respectively.}
\label{P_R3VsRbfinal}
\end{figure*}

The second method we use to evaluate variability involves examining 
the ratio of spectral components at the two different epochs. 
This is useful because it provides a different way to assess our results. 
To evaluate this we consider
$R_{\Ly}=F_{\Ly,HSN}/F_{\Ly,LSN}$
versus $R_{c1300}=F_{c1300,HSN}/F_{c1300,LSN}$
and $R_{CIV}=F_{CIV,HSN}/F_{CIV,LSN}$
versus $R_{c1300}$
in Figure~\ref{P_R3VsRbfinal} (top panel and middle panel, respectively). 
We also plot $R_{\Ly}$ versus $R_{CIV}$ in Figure~\ref{P_R3VsRbfinal} (bottom panel). 
The data shown in red in these figures
are the same data originally plotted in red in 
Figures~\ref{P_LyACIVVsLyAc_bfinal} and \ref{P_LyAVsCIV_bfinal}, i.e., 
objects showing absolute flux variations in the \Ly\ BEL at a level of significance
$> 5\sigma$.  
Of course, when $R$ values are unity it represents no flux change and when, for example, 
$R_{\Ly} = R_{c1300}$ it indicates that the \Ly\ BEL and
the nearby continuum experienced the same fractional increase or decrease in
flux (i.e., grey scaling).  
The results generally confirm the findings of Section~\ref{subsec:corrF}. After taking into account the variances of the 
flux ratios, the correlation coefficients are: 
0.54 for $R_{\Ly}$ versus $R_{c1300}$ (Figure~\ref{P_R3VsRbfinal}, top panel), 
0.69 for $R_{CIV}$ versus $R_{c1300}$ (Figure~\ref{P_R3VsRbfinal}, middle panel), 
and 0.76 for $R_{\Ly}$ versus $R_{CIV}$ (Figure~\ref{P_R3VsRbfinal}, bottom panel). 

\subsection{Time Variations}\label{sec:time}
\begin{figure*}
\centering
\includegraphics[width=0.8\textwidth,angle=0]{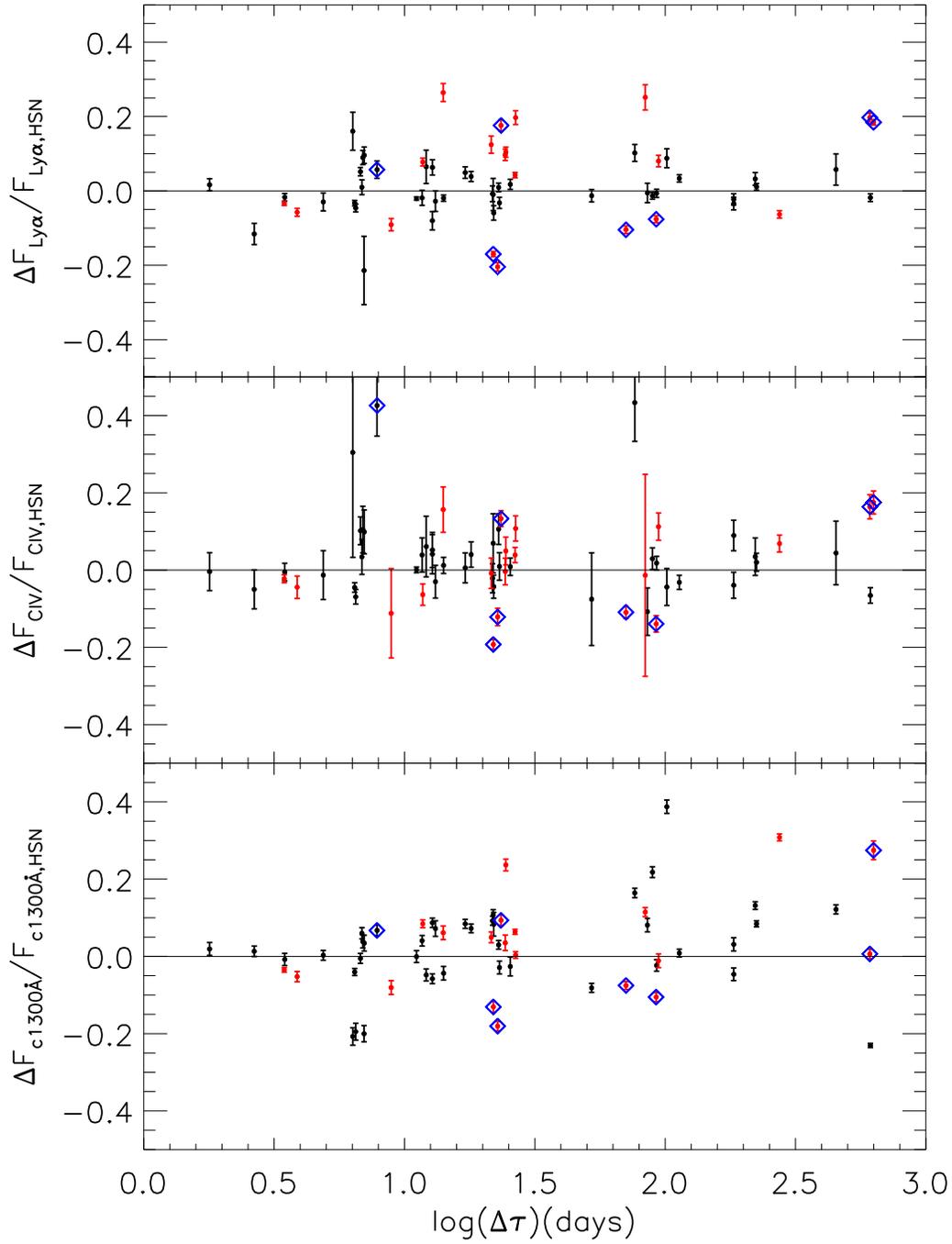}
\caption{Top: Fractional change in \Ly\ BEL flux as a function of time 
between observation epochs in the quasar rest frame on a logarithmic scale, $\log(\Delta\tau)$. 
Middle: Fractional \CIV\ BEL flux changes vs. $\log(\Delta\tau)$. 
Bottom: Fractional continuum changes vs. $\log(\Delta\tau)$.} 
\label{P_F3vsTau_bfinal}
\end{figure*}

For our sample of 61 high-luminosity, high-redshift quasars, the top, middle and bottom panels
of Figure~\ref{P_F3vsTau_bfinal}  show results on the fractional changes 
in \Ly\ BEL flux, \CIV\ BEL flux and continuum flux, respectively, as 
a function of time between observation epochs in the quasar rest frame, $\Delta\tau$. 

\subsection{Examples of Detected Variations}\label{sec:eg}
From the 20 quasars with $>5\sigma$ \Ly\ BEL variations shown in 
Figure~\ref{P_LyACIVVsLyAc_bfinal} (red points), 
we select four representative cases to illustrate. 
Table~\ref{table2} summarizes the two-epoch flux measurements of the BELs and continua for these four quasars, 
along with their associated emission redshifts, rest-frame time interval 
$\Delta\tau$ between the two observation epochs, 
r-band magnitudes, and Galactic extinction-corrected 
\citep{sf11} luminosities at rest-frame 1450\AA. This information for
all 61 quasars can be found in an expanded version of Table 2, which is available in
the electronic version of the journal.

\begin{figure*}
\centering
\includegraphics[width=0.8\textwidth,angle=180]{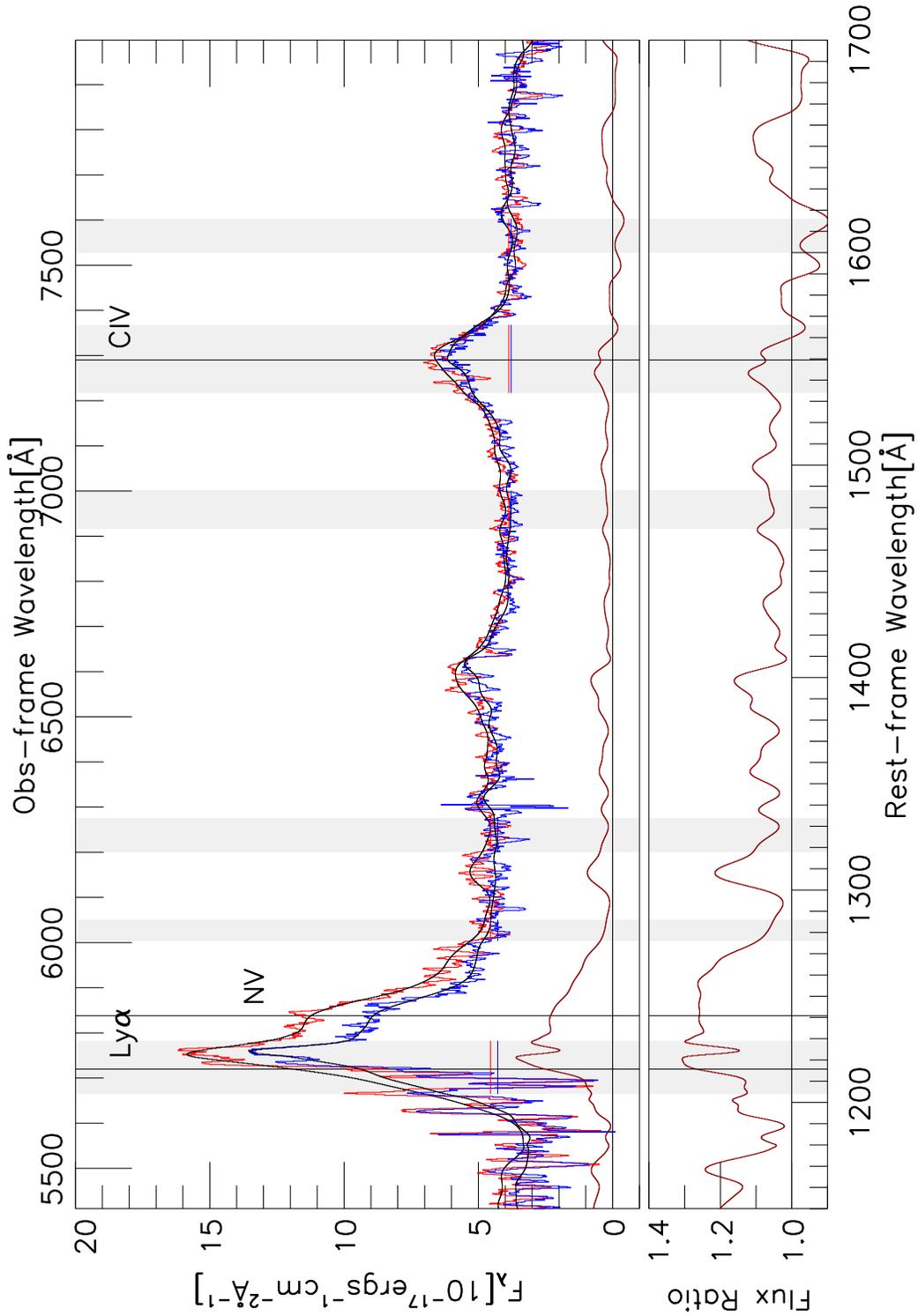}
\caption{(1) Top: Spectra of SDSS J004240.65+141529.6 
at the first HSN-epoch (red) and the second LSN-epoch (blue),
and their difference (dark red). Bottom: ratio spectrum.}
\label{SpAl057}
\end{figure*}

\begin{figure*}
\centering
\includegraphics[width=0.8\textwidth,angle=180]{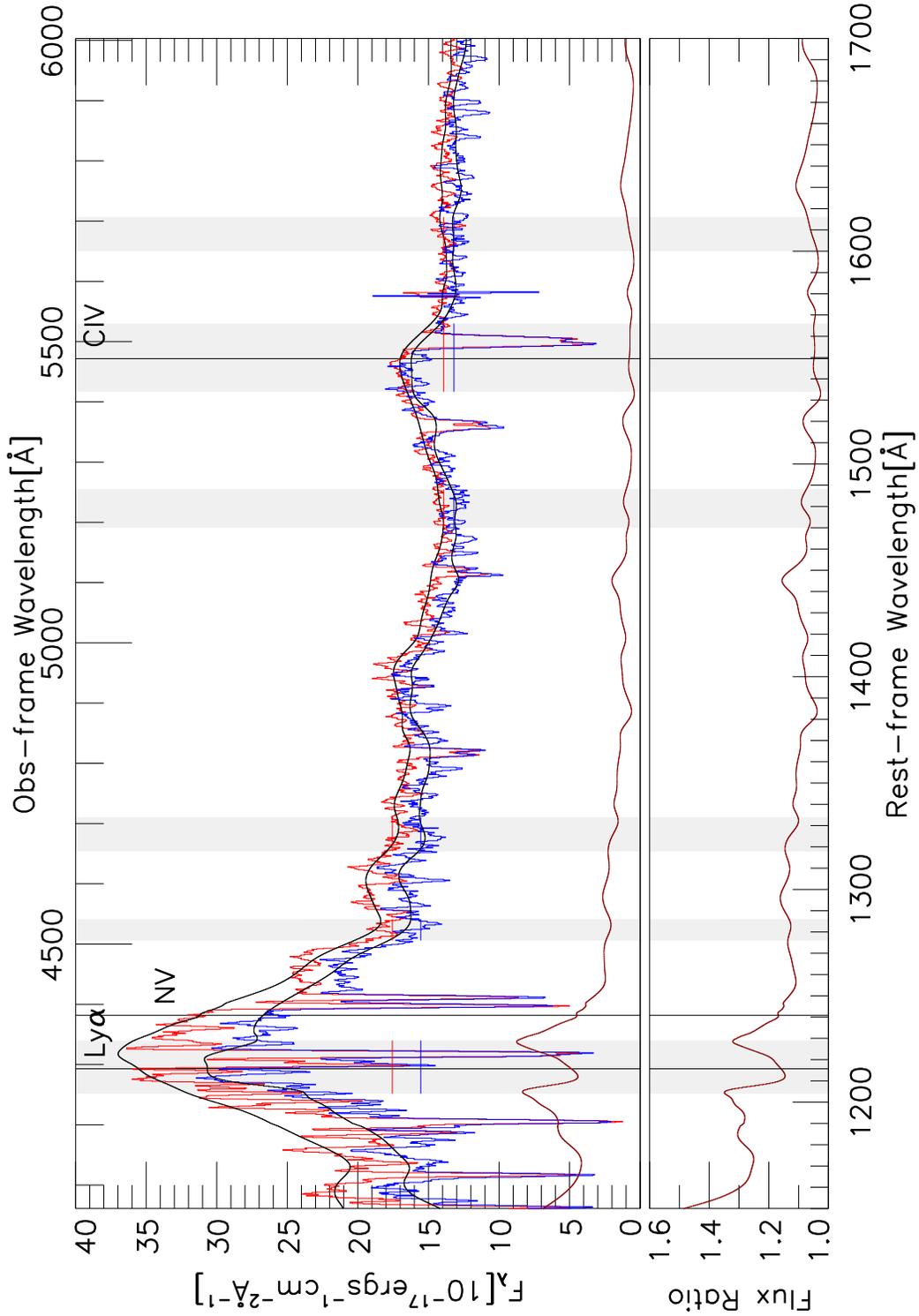}
\caption{(2) Top: Spectra of SDSS J132750.44+001156.8 
at the first LSN-epoch (blue) and the second HSN-epoch (red), 
and their difference (dark red). Bottom: ratio spectrum.}
\label{SpAl013}
\end{figure*}

\begin{figure*}
\centering
\includegraphics[width=0.8\textwidth,angle=180]{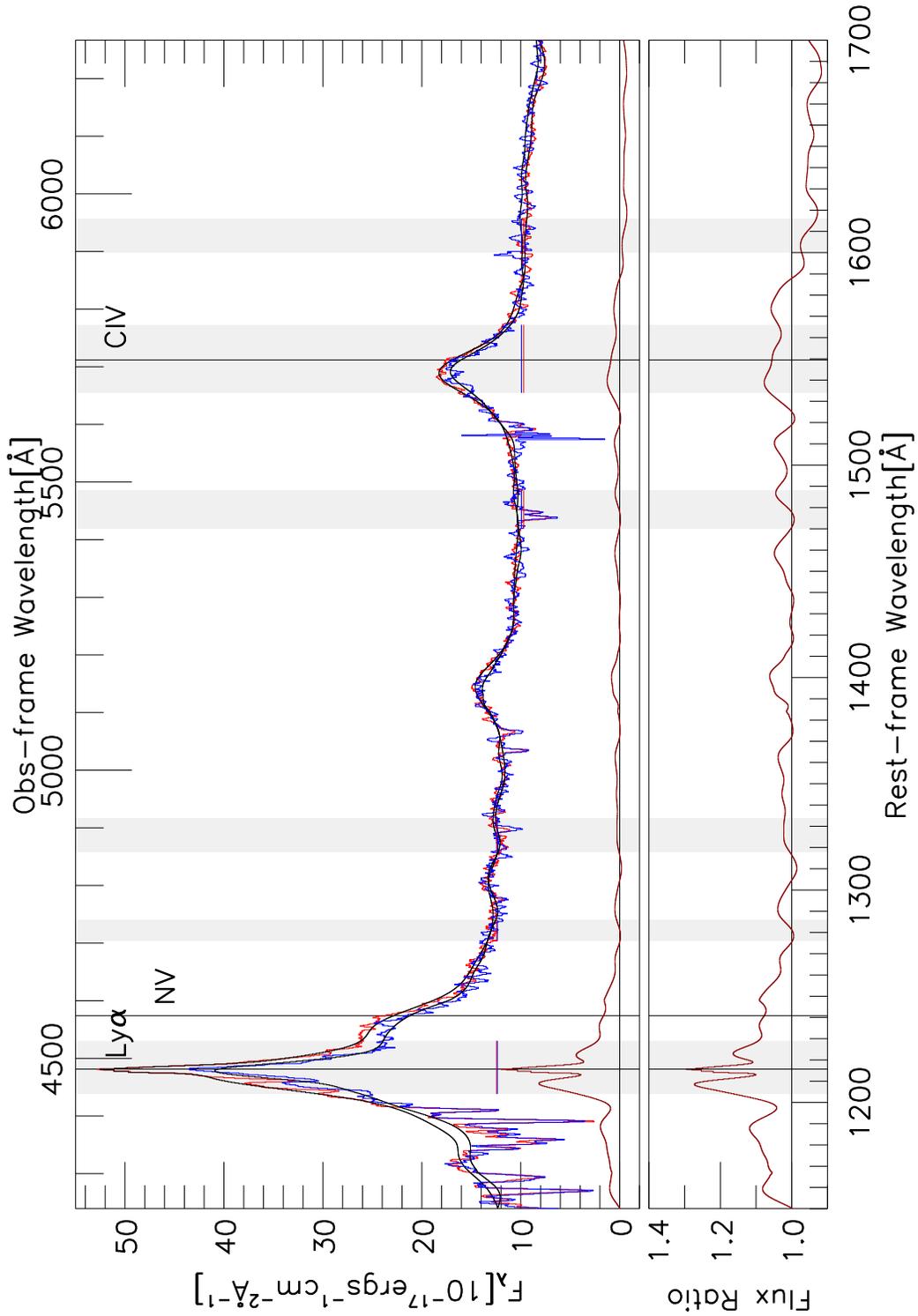}
\caption{(3) Top: Spectra of SDSS J095106.32+541149.8 
at the first LSN-epoch (blue) and the second HSN-epoch (red),
and their difference (dark red). Bottom: ratio spectrum.}
\label{SpAl167}
\end{figure*}

\begin{figure*}
\centering
\includegraphics[width=0.8\textwidth,angle=180]{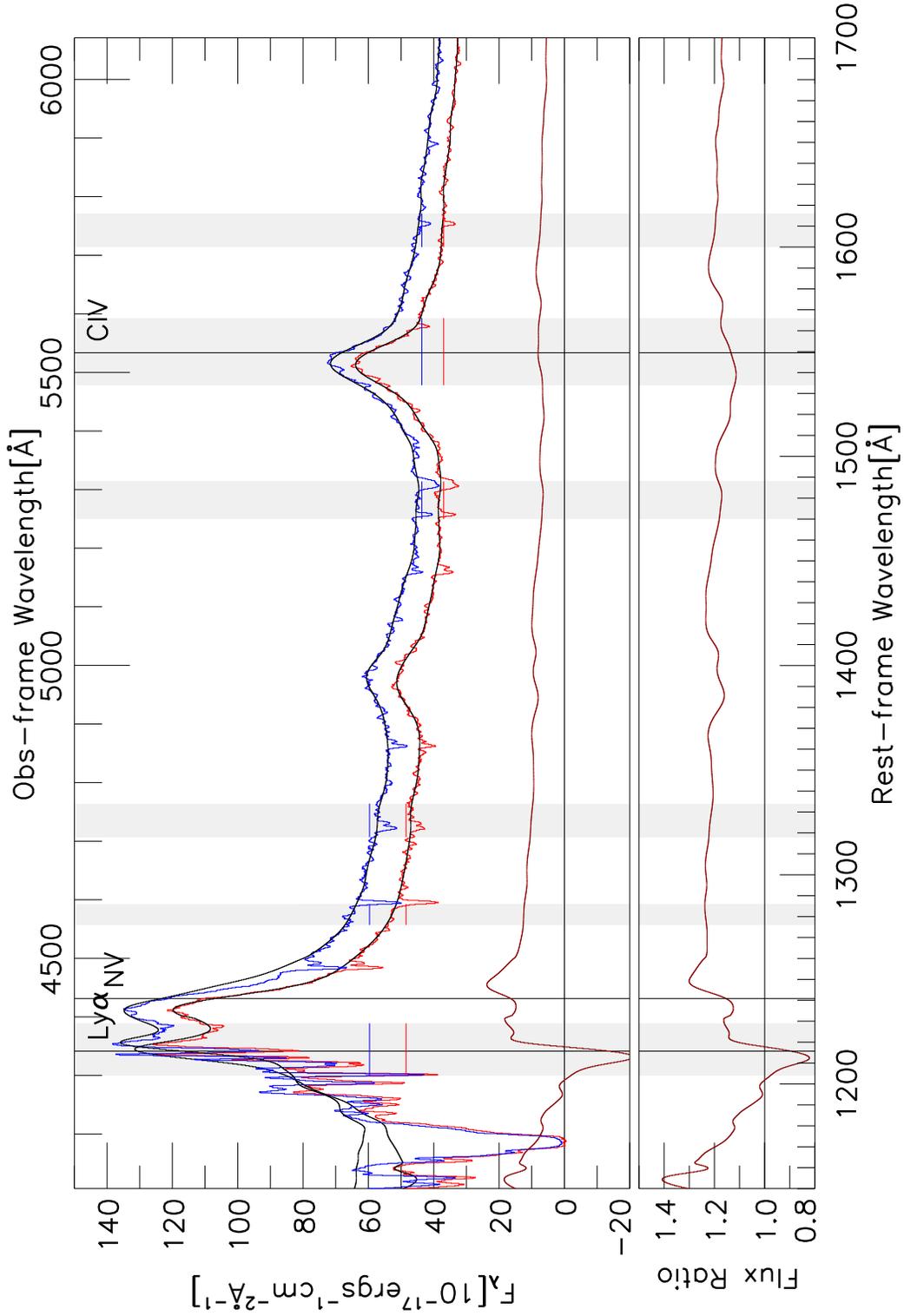}
\caption{(4) Top: Spectra of SDSS J120802.64+630328.9 
at the first HSN-epoch (red) and the second LSN-epoch (blue),
and their difference (dark red). Bottom: ratio spectrum.}
\label{SpAl188}
\end{figure*}

The spectral changes in the four quasars are shown in Figures~\ref{SpAl057}$-$\ref{SpAl188}. 
The red and blue curves in the figures represent the spectra at HSN epochs 
and LSN epochs, respectively. The overplotted black solid lines are the spline fits to the continua and BELs. 
The dark red curve in the top panel of the figures is the flux difference 
between the two epochs, whereas the dark red curve in the bottom panel is the flux ratio between the two epochs.
A brief discussion of the four examples follows. 
  
\subsubsection{SDSS J004240.65+141529.6}
\label{sec:eg_no1}
The spectra of SDSS J004240.65+141529.6 
($z\approx3.70$) are shown in Figure~\ref{SpAl057}, $\sim14$ days apart in the quasar rest frame. 
The redshift of the \Ly\ BEL is somewhat larger than indicated by
the SDSS pipeline redshift, but the redshift of the \CIV\ BEL is a close match. The measured
\Ly\ BEL flux seen in the second LSN-epoch spectrum is $\sim26$\%
fainter than in the first HSN-epoch spectrum, whereas the nearby continuum
flux decreases only $\sim6\%$. 
The relatively fast change in \Ly\ BEL flux in this quasar might be considered 
surprising given its high luminosity.
For example, if the luminosity of this quasar is compared to the luminosity 
of AGN NGC 5548, we find: $(L_{quasar}/L_{AGN})^{0.5} \sim 13.6 $.
Of course, time lags can not be derived from two-epoch data. 
However, if BELRs are larger in luminous quasars, one might naively 
expect fractional \Ly\ BEL flux variations to be slower in high-luminosity objects. Yet the 
fractional variations seen in this quasar are comparable to the ones seen in NGC 5548 \citep{pet94}. 
The observed fast change suggests that a significant fraction of the \Ly\ BELR has a correspondingly 
small size scale along our sightline. For example, this would occur if the \Ly\ BELR were confined to 
a disc-like region, and we were viewing the disc close to face-on. In such a geometry the light travel
time across the BELR along our sightline would be $\sim \sin(i)D_{\Ly\ BELR}$, where
$i$ is the inclination angle relative to the disc's polar axis and 
$D_{\Ly\ BELR}$ is the diameter of the \Ly\ BELR disc. Thus, this observation
would support the idea that the \Ly\ BELR is due to disc emission.  
Alternatively, these short time-scale variations might indicate anisotropic emission from the BELR,
as has been discussed by \citet{ppm89}. 

\subsubsection{SDSS J132750.44+001156.8}
SDSS J132750.44+001156.8 (Figure~\ref{SpAl013}, $z=2.53$) shows increases 
in the \Ly\ BEL flux and nearby continuum flux of $\sim34$ and $\sim12$per
cent during a 
period of $\sim$84 days in the quasar rest frame.
The BELs are very broad, 
such that the wings of the \Ly\ and \CIV\ BELs are not included in the $\sim6200$ km s$^{-1}$ interval
that we measure. Relative to the SDSS pipeline redshift, the redshift of the peak of the \Ly\ BEL appears higher and 
the redshift of the peak of the \CIV\ BEL appears lower.  
A significant flux change in the \CIV\ BEL is not observed.  
The continuum spectral slope at the brighter HSN-epoch is bluer 
than observed at the LSN-epoch.  

\subsubsection{SDSS J095106.32+541149.8}
\label{sec:eg_no3}
The spectra of SDSS J095106.32+541149.8 (Figure~\ref{SpAl167}, $z=2.69$) were taken
$\sim610$ days apart in the quasar rest frame. 
Significant flux changes in both the \Ly\ and \CIV\ BELs are present, whereas 
no significant change in the nearby continuum flux is observed. 
Relative to the initial LSN-spectrum the \Ly\ BEL flux increased $\sim25$\%, the nearby 
continuum flux increased $\sim1$\%, and the \CIV\ BEL flux increased $\sim20$\%.
The redshift of the \Ly\ BEL is consistent with the SDSS redshift, but the redshift
of the \CIV\ BEL is somewhat lower. 
The \Ly\ BEL may show a larger flux change in the line core than in the wing. 

\subsubsection{SDSS J120802.64+630328.9}
The spectra of SDSS J120802.64+630328.9 (Figure~\ref{SpAl188},  $z=2.57$) 
were taken $\sim613$ days apart in the quasar rest frame. No significant flux 
changes in both the \Ly\ and \CIV\ BELs are observed, whereas a significant 
change in the nearby continuum flux is present.  
Relative to the initial HSN-spectrum the \Ly\ BEL flux increased $\sim2$\%, the nearby
continuum flux increased $\sim23$\%, and the \CIV\ BEL flux decreased $\sim7$\%. 
The redshift of the \Ly\ BEL is consistent with the SDSS redshift, but the redshift
of the \CIV\ and \NV\ BEL is somewhat lower. A strong damped \Ly\ (DLA) 
absorption-line system, which is intervening and unrelated to the quasar, is 
identified at a redshift of $z_{abs} = 2.442$.

\subsection{Correlations and their Dependence on Time Interval and Luminosity}

\begin{figure*}
\centering
\includegraphics[width=0.8\textwidth,angle=0]{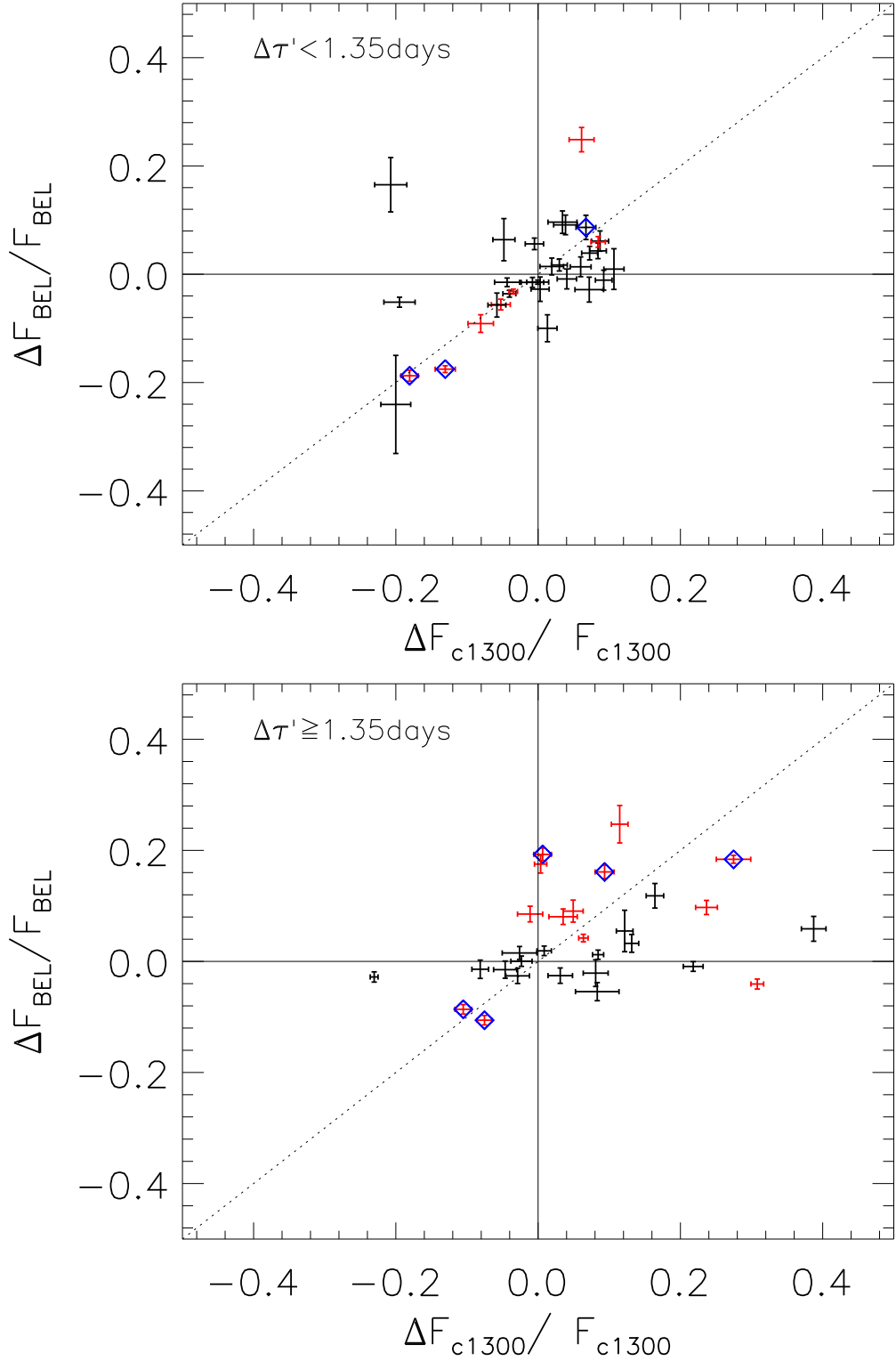}
\caption{Top: Fractional change in BEL flux vs. fractional change in continuum 
between two epochs for 32 quasars with $\Delta\tau^{\prime}<$ 1.35 days. 
Bottom: same as above for 29 quasars with $\Delta\tau^{\prime}\geq$ 1.35
days. See Section 3.5 for the definition of $\Delta\tau^{\prime}$.}
\label{P_BELvsLyAcTaup_bfinal}
\end{figure*}

In order to study any correlations that may depend on the time interval between observations and the quasar 
luminosity, we have formed a new parameter $\Delta\tau^{\prime} = \Delta\tau/(L_{quasar}/L_{AGN})^{0.5}$.
If the variation time increases in proportion to $(L_{quasar}/L_{AGN})^{0.5}$, $\Delta\tau^{\prime}$ would scale the
time intervals to take such an effect into account.  We have used $L_{AGN} = 4.2 \times 10^{43}$ ergs s$^{-1}$, 
which is the luminosity of NGC 5548 at 1300 \AA.
We then divide our sample in half according to $\Delta\tau^{\prime}$. Half of our sample has
$\Delta\tau^{\prime} < 1.35$ days and half has $\Delta\tau^{\prime} > 1.35$ days. Since we found that changes
in the \Ly\ and \CIV\ BELs are well correlated, we have also averaged the individual change results on \Ly\ and \CIV\ to
form a variance-weighted average $\Delta F_{BEL}$. The correlation between BEL flux changes versus
continuum changes is then considered for our sample for short ($\Delta\tau^{\prime}<1.35$ days) and long
 ($\Delta\tau^{\prime}>1.35$ days) luminosity-scaled, rest-frame time intervals.
The results are shown in Figure~\ref{P_BELvsLyAcTaup_bfinal}.  
We find that there is a strong correlation between BEL flux change and continuum change when 
$\Delta\tau^{\prime} < 1.35$ days (correlation coefficient of 0.74), but the correlation is much weaker
when  $\Delta\tau^{\prime} > 1.35$ days (correlation coefficient of 0.29). 
Thus, the overall trend seen in our sample is for BEL and continuum variations to become less correlated for 
the longer luminosity-scaled, rest-frame time intervals. We found this interesting, 
but the origin of this effect is unclear.

\section{Summary and Discussion}
\label{sec:concl}
We have investigated the variability of the \Ly\ BEL in a sample of  
61 quasars observed at two epochs by the SDSS. The quasars
have redshifts that lie in the interval $z=[2.5, 4.3]$ 
and luminosities $3.4 \times
10^{45} \lesssim  \lambda L_{\lambda} \lesssim 3.4 \times 10^{46}$ ${\rm erg\,s^{-1}}$ 
at 1450 \AA.
Variability of the \Ly\ BEL is ubiquitous in our sample on rest-frame 
time-scales ranging from days to years, with $\sim33$ per cent 
of the sample showing \Ly\ BEL variations $>5\sigma$. Thus, despite some earlier
work which reported no significant \Ly\ BEL flux variations 
in a somewhat higher luminosity sample than ours, variability is clearly
present and readily observable in the luminous sample of high-redshift quasars
studied here.

However, during the work that led us to consider our final sample of 61 quasars, examination
of the SDSS sub-exposures led us to conclude that faint ($r>19.7$ mag) and low
S/N ($<$10) spectra may produce unreliable results. 

In addition to measuring variations in the \Ly\ BEL, we also measured the corresponding
continuum and \CIV\ BEL variations. 
Even with two-epoch data, some interesting trends are seen in our sample.
For example, there is not a strong correlation
between \Ly\ BEL variability and nearby continuum variability, and this is consistent with the 
presence of a time lag between the variations. However, 
there is a strong correlation between \Ly\ BEL
variability and \CIV\ BEL variability, and this suggests that these BELRs are at similar distances
from the central ionizing source. 

Finally, some interesting examples are emphasized in the analysis. For example,
in one case (SDSS J004240.65+141529.6) the flux of an \Ly\ BEL increased by
$\sim26$ per cent in only 14 days in the quasar rest frame. This probably indicates that 
a significant fraction of the \Ly\ BELR has a correspondingly 
small size scale along our sightline. In the absence of anisotropic \Ly\ emission from the
BELR, the most likely reason for this fast change would be that the \Ly\ BELR 
has the shape of a disc, and we are viewing the disc close to face-on (Section~\ref{sec:eg_no1}). 
In such a geometry the light travel
time across the BELR along our sightline would be $\sim \sin(i)D_{\Ly\ BELR}$, where
$i$ is the inclination angle relative to the disc's polar axis and 
$D_{\Ly\ BELR}$ is the diameter of the \Ly\ BELR disc. Thus, for example, this observation
appears to be consistent with models where the BELR primarily originates in the chromosphere of
a quasar's face-on accretion disc \citep{mcg+95, mc97}.  
If this is the case, such observations may some day provide constraints on quasar inclination angles, or 
at least allow us to identify quasars that have nearly face-on accretion discs. 
In any case, this work demonstrates that future campaigns of 
spectrophotometric monitoring can improve our understanding 
of the structure of the BELRs of high-luminosity, high-redshift quasars.

\section{Acknowledgement}

We thank Ching-Wa Yip for providing the wavelength-dependent flux ratio 
corrections for our sample and all private discussions and Daniel Vanden Berk 
for some early discussions about variability analysis. 
We also acknowledge the referee for careful reading and useful comments. 

Funding for the SDSS and SDSS-II has been provided by the Alfred P. Sloan
Foundation, the Participating Institutions, the National Science Foundation,
the U.S. Department of Energy, the National Aeronautics and Space
Administration, the Japanese Monbukagakusho, the Max Planck Society, and the
Higher Education Funding Council for England. The SDSS website is http://www.sdss.org/.

The SDSS is managed by the Astrophysical Research Consortium for the Participating Institutions. The Participating Institutions are the American Museum of Natural History, Astrophysical Institute Potsdam, University of Basel, University of Cambridge, Case Western Reserve University, University of Chicago, Drexel University, Fermilab, the Institute for Advanced Study, the Japan Participation Group, Johns Hopkins University, the Joint Institute for Nuclear Astrophysics, the Kavli Institute for Particle Astrophysics and Cosmology, the Korean Scientist Group, the Chinese Academy of Sciences (LAMOST), Los Alamos National Laboratory, the Max-Planck-Institute for Astronomy (MPIA), the Max-Planck-Institute for Astrophysics (MPA), New Mexico State University, Ohio State University, University of Pittsburgh, University of Portsmouth, Princeton University, the United States Naval Observatory, and the University of Washington.

This research has made use of the NASA/IPAC Extragalactic Database (NED) which 
is operated by the Jet Propulsion Laboratory, California Institute of 
Technology, under contract with the National Aeronautics and Space 
Administration. 

\bibliography{BALref}

\clearpage

\begin{deluxetable}{lcccccc}
\tablecaption{Flux ratios for four quasars and surrounding compact objects
before and after the grey recalibration correction. 
\label{table1}}
\tablewidth{0pt}
\tablehead{\colhead{} & \multicolumn{2}{c}{Quasars} & \multicolumn{4}{c}{Compact Objects} \\
\cline{1-7}\\
\colhead{} & \colhead{SDSS J} & \colhead{$C^{\rm before}_{\rm band}$}\tablenotemark{1} &
\colhead{SDSS J} & \colhead{$C^{\rm before}_{\rm band}$}\tablenotemark{1} & 
\colhead{$\overline{C}^{\rm before}_{\rm band}$}\tablenotemark{2} &
\colhead{$\overline{C}^{\rm after}_{\rm band}$}\tablenotemark{3} }
\startdata

1 &004240.65+141529.6 & 0.908 & 003408.61+134839.0 & 0.986 & 0.984$\pm$0.015  & 1.000$\pm$0.015 \\
 & & & 003448.91+134439.3 & 0.955 && \\
 & & & 003441.35+153517.8 & 1.055 && \\
 & & & 003651.49+145035.9 & 0.986 && \\
 & & & 003854.02+143742.9 & 0.969 && \\
 & & & 003707.21+155552.5 & 0.951 && \\
\cline{1-7}
2 &132750.44+001156.8 & 1.286 & 132456.66$-$010534.6 & 1.185 & 1.177$\pm$0.012 & 1.000$\pm$ 0.010 \\
 & & & 132304.59+000924.5 & 1.119 && \\
 & & & 132634.99$-$011108.2 & 1.127 && \\
 & & & 132730.91$-$011024.8 & 1.218 && \\
 & & & 132404.47$-$004428.8 & 1.179 && \\
 & & & 132510.57$-$003233.6 & 1.202 && \\
 & & & 133136.77+001959.3 & 1.206 && \\
\cline{1-7}
3 & 095106.32+541149.8 & 1.188 & 094420.28+540810.2 & 1.193 & 1.172$\pm$0.029& 1.000$\pm$0.025 \\
 & & & 094659.80+544546.7 & 1.191 && \\
 & & & 094920.08+543803.2 & 1.030 && \\
 & & & 095003.13+541821.7 & 1.217 && \\
 & & & 095253.45+550957.7 & 1.231 && \\
\cline{1-7}
4 & 120802.64+630328.9 & 1.264 & 115510.14+624653.5 & 1.048 & 1.047$\pm$0.002 & 1.000$\pm$0.002\\
 & & & 120039.52+623428.5 & 1.113 && \\
 & & & 121642.58+623405.9 & 0.978 && \\
 & & & 121425.49+623009.0 & 1.050 && \\
\cline{1-7}
\enddata
\tablecomments{We use the rest-frame wavelength bands at
1260--1285 \AA, 1318--1380 \AA, 1430--1525 \AA\ and 1580--1620 \AA,
ensuring no prominent emission lines at the chosen regions in four quasars.} 
\tablenotetext{1}{Flux ratio $C^{before}_{band}$ of each object 
before recalibration using the method of \citet{ycv+09}.}
\tablenotetext{2}{Mean of $C^{before}_{band}$ of the compact objects in each group
before recalibration using the method of \citet{ycv+09}.}
\tablenotetext{3}{Mean of $C^{after}_{band}$ of the compact objects in each group
after applying the recalibration.}
\end{deluxetable}

\clearpage
\begin{deluxetable}{lcccccccccccccccc}
\tablecaption{BEL and continuum flux measurements for four quasars after recalibration using surrounding compact objects 
\label{table2}} 
\tablewidth{0pt} 
\tabletypesize{\tiny} \rotate 
\tablehead{\colhead{} & \colhead{} & \multicolumn{2}{c}{MJD} & \colhead{} &
\colhead{$\Delta\tau$} & \colhead{} & \colhead{$\lambda L_{1450\AA}$} & \multicolumn{4}{c}{HSN-epoch} & \colhead{} & \multicolumn{4}{c}{LSN-epoch} \\ 
\cline{3-4} \cline{9-12} \cline{14-17} \\ 
\colhead{} & \colhead{SDSS J} & \colhead{HSN} & \colhead{LSN} & \colhead{$z_{em}$} & \colhead{(days)} & \colhead{r} & \colhead{($\rm ergs\,s^{-1}$)} & \colhead{$F_{Ly\alpha}$} & \colhead{$F_{c1300\AA}$} & \colhead{$F_{C \tiny IV}$} & \colhead{$F_{c1600\AA}$} & \colhead{} & \colhead{$F_{Ly\alpha}$} & \colhead{$F_{c1300\AA}$} & \colhead{$F_{C \tiny IV}$} & \colhead{$F_{c1600\AA}$} \\ 
\colhead{} & \colhead{(1)} & \colhead{(2)} & \colhead{(3)} & \colhead{(4)} & \colhead{(5)} & \colhead{(6)} & \colhead{(7)} & \colhead{(8)} & \colhead{(9)} & \colhead{(10)} & \colhead{(11)} & \colhead{} & \colhead{(12)} & \colhead{(13)} & \colhead{(14)} & \colhead{(15)}} \startdata
1 & 004240.65+141529.6 & 51817 & 51884 & 3.70 & 14.08 & 19.20 & 9.31$\times10^{45}$ &
6.15$\pm$0.11 & 4.55$\pm$0.05 & 1.77$\pm$0.08 & 3.87$\pm$0.06 & &
4.52$\pm$0.10 & 4.28$\pm$0.06 & 1.49$\pm$0.07 & 3.78$\pm$0.05 \\ 
2 & 132750.44+001156.8 & 51959 & 51663 & 2.53 & 83.88 & 18.44 & 1.03$\times10^{46}$ &
9.93$\pm$0.23 & 17.56$\pm$0.14 & 0.83$\pm$0.15 & 13.93$\pm$0.10 & &
7.43$\pm$0.23 & 15.55$\pm$0.15 & 0.84$\pm$0.16 & 13.20$\pm$0.11 \\
3 & 095106.32+541149.8 & 54530 & 52282 & 2.69 & 609.65 & 18.89 & 8.45$\times10^{45}$ &
23.28$\pm$0.18 & 12.45$\pm$0.09 & 6.29$\pm$0.11 & 9.71$\pm$0.08 & &
18.69$\pm$0.21 & 12.37$\pm$0.12 & 5.26$\pm$0.16 & 9.93$\pm$0.11 \\
4 & 120802.64+630328.9 & 52337 & 54525 & 2.57 & 612.65 & 17.70 & 3.30$\times10^{45}$ &
46.93$\pm$0.31 & 48.48$\pm$0.17 & 16.59$\pm$0.22 & 36.96$\pm$0.12 & &
47.78$\pm$0.37 & 59.65$\pm$0.20 & 17.68$\pm$0.24 & 43.64$\pm$0.15 \\ 
\enddata \tablecomments{All flux units are in $10^{-17}$\,erg\,s$^{-1}$\,cm$^{-2}$\,\AA$^{-1}$.
This information for
all 61 quasars can be found in an expanded version of Table 2, which is available in
the electronic version of the journal.}
\end{deluxetable}
\clearpage


\end{document}